\documentclass[letterpaper,10pt]{article} 

\usepackage{opticameet3} 

\usepackage{cite}
\usepackage{amsmath}
\usepackage{amssymb}
\usepackage{amsthm}
\usepackage{thmtools}
\usepackage{graphicx}
\usepackage{xcolor}
\usepackage{tikz}
\usepackage{pgfplots}
\usepackage{bbm}
\usepackage{multicol}
\usepackage{caption}
\usepackage[colorlinks=true,bookmarks=false,citecolor=blue,urlcolor=blue]{hyperref} 

\DeclareMathOperator{\coeff}{coeff}

\pgfplotsset{compat=1.18}

\begin{document}

\title{Performance-Complexity-Latency Trade-offs of Concatenated RS-SDBCH Codes}


\author{Alvin Y. Sukmadji and Frank R. Kschischang}

\address{Department of Electrical and Computer Engineering, University of Toronto, Toronto, Ontario, M5S 3G4, Canada}

\email{alvin.sukmadji@mail.utoronto.ca, frank@ece.utoronto.ca} 

\begin{abstract} 
Performance-complexity-latency trade-off curves for rate-0.88 concatenated outer
Reed--Solomon codes and inner Chase-algorithm-based
soft-decision Bose--Ray-Chaudhuri--Hocquenghem
codes with PAM4 constellation using bit-interleaved coded modulation
and multilevel coding coded modulation schemes over the AWGN channel are presented.
\end{abstract}

\section{Introduction} 
Concatenated codes with Reed--Solomon (RS) outer codes and
Bose--Ray-Chaudhuri--Hocquenghem (BCH) inner codes are
attractive
for high-throughput communication systems, such as
data centre interconnects,
that require low complexity and low latency~\cite{yang,wang2023}.
Such coding schemes are often used in conjunction with
four-level pulse amplitude modulation
(PAM4) due to the greater spectral efficiency
compared with two-level non-return-to-zero (NRZ) modulation~\cite{chang,ieee8023df}.
Expanding upon our previous work~\cite{sukmadji2024},
this paper presents the result of a computer search for
combinations of inner and outer codes that achieve
a good trade-off among performance, complexity, and latency.
More specifically,
in this paper, we first describe a semi-analytical
formula to determine the output frame error rate
of concatenated codes with outer RS and inner soft-decision BCH (RS-SDBCH) codes
decoded using the Chase-II algorithm~\cite[Alg.~2]{chase}.
We then use the formula in a code search
to determine the performance-complexity-latency trade-offs of concatenated
RS-SDBCH codes with PAM4 constellation using the bit-interleaved coded modulation
(BICM) and multilevel coding (MLC) coded modulation schemes over the additive
white Gaussian noise (AWGN) channel.



Throughout this paper, we denote by $\mathcal{C}(n,k,t)$ a linear systematic code
with block length $n$, dimension $k$, and error-correcting radius $t$.

\section{Concatenated RS-SDBCH Coding System Description}
\begin{figure}[ht]
\centering
\includegraphics[scale=0.95]{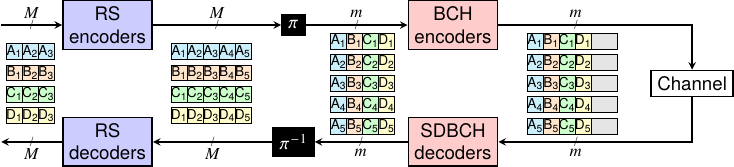}
\captionsetup{font={footnotesize}}
\caption{Example of an RS-SDBCH concatenated coding system with $M=4$ outer RS(5, 3) codes,
over a $B$-bit symbol alphabet and $m=5$ inner BCH($n,4B$) codes.
Here, $L_{i,j}=1$ for all $i\in[M]$, $j\in[m]$.
Each square represents one ($B$-bit) RS symbol.}
\label{fig:system}
\end{figure}

We consider a concatenated system, as shown in Fig.~\ref{fig:system},
consisting of $M$ outer RS($N,K,T$) codes (with $B$ bits
per RS-symbol, where $B$ is assumed---for compatibility with PAM4 modulation---to be even)
concatenated with $m$
inner BCH($n,k,t$) codes, each shortened from a
code of length $2^b-1$. The bits from the outer RS codes
are interleaved into the information positions of the
inner BCH codewords in a symbol-wise manner. Let $L_{i,j}$ denote
the number of RS symbols that are interleaved from the $i$th RS codeword
to the $j$th BCH codeword. For simplicity, we may assume that those
$L_{i,j}$ RS symbols
are placed next to each other, forming a \emph{strip} of $L_{i,j}$ RS symbols.

We consider that the PAM4 constellation is used with BICM and MLC coded modulation schemes,
and each RS symbol comprises $B/2$ PAM4 symbols.
We call the first and second bits of each PAM4 symbol to be the most significant bit (MSB)
and the least significant bit (LSB) respectively.
In the BICM scheme, we assume that $B$ divides $k$ and
a Gray-labeled PAM4 constellation (i.e.,
$00\mapsto -3$, $01\mapsto -1$, $11\mapsto +1$, $10\mapsto +3$) is used.
The $k$ information bits and the $n-k$ parity bits
are mapped into $k/2$ and $\left\lceil\frac{n-k}{2}\right\rceil$
PAM4 symbols, respectively. In the MLC scheme, on the other hand,
only the LSBs of the bits from the interleaved RS symbols are protected by
the inner BCH code, while the MSBs are not protected by an inner code.
We also assume that $B/2$ divides $k$
and a natural-labeled PAM4 constellation
(i.e., $00\mapsto -3$, $01\mapsto -1$, $10\mapsto +1$, $11\mapsto +3$)
is used.
Thus, there are
$2k$ bits ($k$ protected and $k$ not protected by the inner code) formed by $2k/B$ RS symbols
that are mapped into $k$ PAM4 symbols that are transmitted over the channel.
The $n-k$ BCH parity bits
protecting the LSBs are mapped using Gray-labelled PAM4 just like that in BICM,
and $\left\lceil\frac{n-k}{2}\right\rceil$ additional PAM4 symbols are transmitted.

The inner BCH decoder uses the
Chase (or Chase-II)~\cite[Alg.~2]{chase} soft-decision decoder
with $J$ test bits. The Chase decoder will first find the $J$
least reliable positions of the input bits and generate $2^J$ test input patterns.
Then, the decoder will attempt to decode the $2^J$ patterns
and select the output codeword that has the lowest ``analog weight''
with respect to the initial input word.


\section{Estimating the Frame Error Rate}
\label{sec:fer-estimation}
We have shown in~\cite[eq.~(4)]{sukmadji2024} that
the coefficient of $x^uy^v$ of
the bivariate generating function
\vspace{-1ex}
$$W_{B,k,L}(x,y)=\left(1+\left((1+x)^{B}-1\right)y\right)^L(1+x)^{k-BL}$$
is the number of bit-error patterns in a vector of length $k$ bits
with $u$ bit errors that induce $v$ RS-symbol errors in a strip of $L$ RS-symbols.
Since we are using PAM4 modulation scheme, we can generalize
this generating function so that the indeterminate $x$
tracks PAM4-symbol errors rather than bit errors. That is, we can define the
coefficient of $x^uy^v$ of $W_{B/2,k',L}(x,y)$
to be the number of PAM4-symbol-error patterns in a vector of length $k'$ PAM4 symbols
with $u$ PAM4-symbol errors that induce $v$ RS-symbol errors in a strip of $L$ RS symbols.

Let $U_j$ denote the PAM4-symbol-error weight of the information bits
(plus the corresponding bits not protected by the inner code in the case of MLC)
at the output of the $j$th SDBCH decoder.
Let
$V_{i,j}$ denote the number of RS-symbol errors in a strip of length
$L_{i,j}$ RS-symbols produced at the output of the $j$th SDBCH decoder.
In the case of BICM, using the law of total probability, we have
\vspace{-1ex}
\begin{align}
\displaystyle\Pr(V_{i,j}=v)=\sum_{u=0}^{k/2}\Pr(V_{i,j}=v\mid U_j=u)\Pr(U_j=u),~\text{\small where}~
\Pr(V_{i,j}=v\mid U_j=u)=\frac{\coeff_{x^uy^v}(W_{B/2,k/2,L_{i,j}}(x,y))}{\binom{k/2}{u}}
\label{eqn:symb-err-dist-bicm}
\end{align}
\vspace{-1ex}
for all $u\in\left\{0,1,\ldots,\frac{k}{2}\right\}$ and $v\in\left\{0,1,\ldots,L\right\}$.
In the case of MLC, we have
\begin{align}
\displaystyle\Pr(V_{i,j}=v)=\sum_{u=0}^{k}\Pr(V_{i,j}=v\mid U_j=u)\Pr(U_j=u),~\text{\small where}~
\Pr(V_{i,j}=v\mid U_j=u)=\frac{\coeff_{x^uy^v}(W_{B/2,k,L_{i,j}}(x,y))}{\binom{k}{u}}.
\label{eqn:symb-err-dist-mlc}
\end{align}
for all $u\in\left\{0,1,\ldots,k\right\}$ and $v\in\left\{0,1,\ldots,L\right\}$.
We approximate via Monte Carlo simulations the distribution of $U_j$ as it is
difficult to determine this distribution analytically.
The distributions of $U_j$ for different inner code and channel parameters are precomputed
and used when computing
\eqref{eqn:symb-err-dist-bicm} or \eqref{eqn:symb-err-dist-mlc}.

The rest of the derivation is then identical to that in~\cite[eqs.~(7)--(9)]{sukmadji2024}.
The number of RS-symbol errors in the received $i$th RS word is $Y_i=V_{i,0}+\cdots+V_{i,m-1}$.
Since $V_{i,0},\ldots,V_{i,m-1}$ come from independent BCH decoders, they are independent
random variables. Therefore,
the distribution of $Y_i$ is the convolution of the distributions of the $V_{i,j}$'s.
If at least one of the RS word contains more than $T$ RS-symbol errors, then a frame error occurs.
The probability of a frame error occurring is typically difficult to compute
due to the correlation between $Y_0,\ldots,Y_{m-1}$.
Instead, we use a simple union bound to estimate the probability of a frame error, i.e.,
\vspace{-1.8ex}
$$\Pr(\text{frame error}) = \sum_{i=0}^{n-1}\Pr(Y_i>T).$$

\vspace{-1.8ex}
Fig.~\ref{fig:fer} show examples of the output FER obtained using the formula
derived in this section compared with the results of Monte Carlo FER simulation
with the KP4 outer code, i.e., RS(544, 514, 15) with $B=10$ over the AWGN channel.
At low input signal-to-noise ratio (SNR), there are
gaps between the data points obtained from the formula and the simulation.
However, those gaps are closed as the input SNR grows.
Using the formula outlined in this section, we can skip time-consuming Monte Carlo simulations
for the RS-SDBCH system and we can
predict at which input SNR values the codes achieve the target FER.



\section{Performance-Complexity-Latency Trade-offs}
We define the metric of performance as the gap to the constrained Shannon limit (CSL) at $10^{-13}$ post-FEC FER.
We define the metric of complexity as the number of elementary decoding operations for the inner and outer decoders
(integer and real-valued additions and multiplications,
table lookups, comparisons needed for sorting, etc.) per decoded information bit.
We define the metric of latency as the overall block length, i.e., $mn$ and $m(n+k)$ for BICM and MLC, respectively.

\begin{figure}[t]
\begin{minipage}[t]{.45\textwidth}
\centering
\includegraphics[scale=0.99]{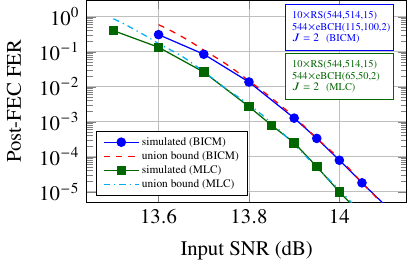}
\vspace{-3mm}
\captionsetup{font={footnotesize}}
\caption{FER versus input SNR for RS-SDBCH codes (with code parameters as indicated)
with PAM4 modulation using BICM and MLC over the AWGN channel.}
\label{fig:fer}
\end{minipage}
\hfill
\begin{minipage}[t]{.535\textwidth}
\centering
\includegraphics[scale=0.99]{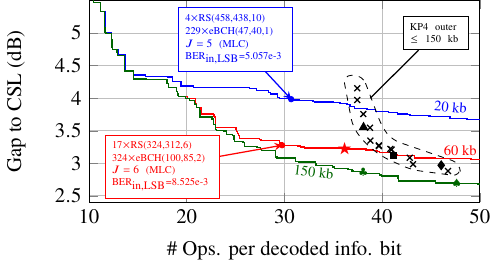}
\vspace{-3mm}
\captionsetup{font={footnotesize}}
\caption{Performance-complexity-latency trade-offs of Pareto-efficient RS-SDBCH codes with rate
$0.88\pm 0.005$. Each curve shows operating points with the maximum latency indicated.}
\label{fig:pareto}
\end{minipage}
\end{figure}

We perform an extensive code search of codes with rate 0.88 $\pm$ 0.005
with outer RS codes (not only KP4) having $B=10$ bits, $T=1,\ldots,20$, and $K=N-2T$,
inner extended $t=1,2,3$ SDBCH codes with $b=5,6,\ldots,11$, $k=n-bt-1$, $J=1,\ldots,6$, and inner rate of $\leq 0.99$,
and maximum latency of 20, 60, and 150 kilobits.
We also assume that $L_{i,j}=\left\lceil N/m\right\rceil$ or $\left\lfloor N/m\right\rfloor$,
i.e., given an RS codeword, the number of RS symbols that are interleaved into each BCH codeword
is about equal.
About 7.5 million code combinations are evaluated using the formula
described in Sec.~\ref{sec:fer-estimation} and
the performance-complexity-latency trade-off curves of the Pareto-efficient codes are shown
in Fig.~\ref{fig:pareto} along with the parameters of some codes on the
Pareto frontier; see \cite{sukmadji-tables} for a complete listing. Also included in Fig.~\ref{fig:pareto} are the best
combinations of outer KP4 and inner SDBCH that our search could find.
These results show
that KP4 is not the best choice of outer code as there are
other outer RS codes leading to either better performance
at the same complexity or lower complexity at the same performance,
such as those listed in Table~\ref{tab:rate-88-kp4}.

We observe that all codes on the Pareto frontier use MLC instead of BICM.
This is due to the fact that only half of the bits from the outer codes, namely the LSBs, are
encoded and decoded by the inner codes. Thus, the number of inner decoding operations are effectively
halved compared with BICM. For a fixed inner rate, the length of the eBCH code
used in MLC is roughly half of the block length used in BICM,
while the number of parity bits is about the same.
Thus, the BCH codes that are used to decode the LSBs in the MLC scheme have
lower rate, and therefore generally have stronger error-correcting performance.
Furthermore, given a corrected LSB,
conditionally demapping the corresponding MSB rarely results in an incorrect bit.


We should note that the conclusions that we have drawn in this paper are based
on the code search space that we have defined as well as our choice of figures of merit.
Enlarging the search space,
for example, by allowing $B=8$ or $B=12$ (which, to allow fair comparisons with KP4,
we did not consider) may change the Pareto frontier.
The tools that we have described in this paper readily extend to such enlarged search
spaces.

\vspace{-0.2ex}
\begin{table}[t]
\centering
\caption{Parameters of some selected Pareto-optimal rate-0.88 codes with KP4 and other RS outer codes}
\vspace{-3mm}
\scalebox{0.795}{
\renewcommand{\tabcolsep}{2pt}
\begin{tabular}{|c|c|c|c|c|c|c|c|c|c|c||c|c|c|c|c|c|c|c|c|c|c|c|c|}\hline
\multicolumn{11}{|c||}{KP4 outer code ($N=544$, $B=10$, $T=15$)} & \multicolumn{13}{|c|}{Other RS outer codes ($B=10$)}\\\hline
$M$ & $m$ & $n$ & $b$ & $t$ & $J$ & Lat. & Compl. & Gap (dB) & BER${}_\text{in,LSB}$ & Type & $M$ & $N$ & $T$ & $m$ & $n$ & $b$ & $t$ & $J$ & Lat. & Compl. & Gap (dB) & BER${}_\text{in,LSB}$ & Type\\\hline
${}^\blacktriangle$10 & 544 & 57 & 6 & 1 & 2 & 58208 & 38.10 & 3.556 & 6.945e-3 & MLC & {\color{red}${}^\bigstar$}8 & 689 & 14 & 689 & 47 & 6 & 1 & 5 & 59943 & 36.22 & 3.220 & 8.947e-3 & MLC\\
${}^\blacksquare$22 & 544 & 125 & 7 & 2 & 4 & 127840 & 41.14 & 3.121 & 9.743e-3 & MLC & {\color{green!40!black}${}^\clubsuit$}15 & 752 & 10 & 752 & 90 & 7 & 2 & 6 & 124080 & 38.07 & 2.857 & 1.189e-2 & MLC\\
${}^\blacklozenge$25 & 544 & 142 & 8 & 2 & 6 & 145248 & 46.05 & 2.968 & 1.096e-2 & MLC & {\color{green!40!black}${}^\spadesuit$}16 & 857 & 15 & 857 & 95 & 7 & 2 & 6 & 149975 & 47.65 & 2.692 & 1.402e-2 & MLC\\\hline
\end{tabular}}
\label{tab:rate-88-kp4}
\end{table}
\end{document}